\documentclass[prd,twocolumn,nofootinbib]{revtex4}

\usepackage{dblfloatfix}
\usepackage{dcolumn}
\usepackage{graphicx}
\usepackage{epsfig} 
\usepackage{amsmath}
\usepackage{amsfonts}
\usepackage{amssymb}
\usepackage{color}
\usepackage{float}
\usepackage[caption = false]{subfig}
\usepackage{natbib}	

\begin{document}

\title{New constraints on the linear growth rate using cosmic voids in the SDSS DR12 datasets }

\author{I. Achitouv$^{\dagger\dotplus}$}

\affiliation{$^\dagger$Laboratoire Univers et Th\'eories (LUTh), UMR 8102 CNRS, Observatoire de Paris, Universit\'e Paris Diderot, 5 Place Jules Janssen, 92190 Meudon, France\\
$^\dotplus$APC, Univ Paris Diderot, CNRS/IN2P3, CEA/lrfu, Obs de Paris, Sorbonne Paris Cité, France\\\\}

\begin{abstract}
We present a new analysis of the inferred growth
rate of cosmic structure measured around voids, using the LOWZ and the CMASS samples in the twelfth data release (DR12) of SDSS.  Using a simple multipole analysis we recover a value consistent with $\Lambda$CDM for the inferred linear growth rate normalized by the linear bias: the $\beta$ parameter. This is true in both the mock catalogues and the data, where we find $\beta=0.33\pm0.06$ for the LOWZ sample and $\beta=0.36\pm0.05$ for the CMASS sample. This work demonstrates that we can expect redshift-space distortions around voids to provide unbiased and accurate constraints on the growth rate, complementary to galaxy clustering, using simple linear modelling. 
\end{abstract}

\maketitle

\section{Introduction}
The growth rate of cosmic structure $f$ tells us how fast density fluctuations $\Delta$ grow with respect to the scale factor of the Universe $a$: 
\begin{equation}
f\equiv \frac{d\ln \Delta}{d\ln a}\label{fdef}
\end{equation}

\noindent Its measurement as a function of time and scale is a key cosmological probe, very sensitive to the nature of gravity (e.g. \cite{Linder,Huterer}). To infer the growth rate, we can measure redshift-space distortions (RSD) in the galaxy clustering signal. These distortions are due to the peculiar motions of galaxies, which on large scales have a coherent motion sourced by the gravitational potential of cosmic structures. This gravitational potential is itself proportional to the growth rate, in the linear regime. For standard General Relativity (GR) and isotropic cosmologies, the linear growth rate does not depend on the comoving spatial scale \cite{Peebles} and can be approximated by $f \sim \Omega_m(z)^\gamma$ where $\Omega_m$ is the matter density parameter at redshift $z$, and $\gamma$ is a constant. For a $\Lambda$CDM Universe $\gamma \sim 0.55$ \cite{Linder,Huterer}, independently of the environment.  Constraints on the linear growth rate made with galaxy-galaxy correlation function measurements in redshift-space are well known, e.g.~\citep{Peacock2001,Tegmark2006,Blake2011,Reid2012,delaTorre2013,Beutler6dF} . These measurements have shown a general consistency with the $\Lambda$CDM
cosmological model, up to a $2.5\%$ precision, albeit in some cases
showing tension with the predictions of the latest Cosmic Microwave
Background measurements \cite{Planck}. 

\medskip
\noindent On the other hand, it was only recently that the growth rate has been inferred using the RSD pattern around cosmic voids. There are at least two reasons to perform this consistency test of the linear growth rate. First, certain models of modified gravity, such as $f(R)$ \cite{Hu_Sawicki_2007}, rely on the 
the chameleon screening mechanism  \citep{Khoury2004} which suppresses the 5th force in high density regions, while in under-dense regions the total gravitational force is enhanced (due to the presence of the 5th force), resulting in specific imprints on void abundance and density profiles around underdense regions (e.g. \cite{ANP,Achitouv16,ABPW,Clampitt2012,Zivicketal2015,Cai15}). These theories would naturally lead to an environmentally-dependent growth rate. In fact, in the non-linear regime, the linear growth rate is also sensitive to the underlying density, as shown in \cite{AchitouvCai}. For very large under-dense regions, the effective cosmological parameters are expected to be different to the
globally-averaged parameters, but the quantification of this critical
scale can also serve as an interesting test for departures from
Einstein gravity. Second, the formation and evolution of cosmic voids is non-linear and reduced compared to the dynamics of dark matter halos or the evolution of overdense regions with $\Delta(r)\gg 1$. This is why we can expect that quasi-linear or linear models can describe the RSD around voids relatively well, although recent works have shown the limitation of this assumption \cite{Cai16,Achitouv17,Nadathur17, AchitouvCai}. The first studies that have tested the growth rate measurements using RSD around cosmic voids in galaxy surveys, have used a Gaussian Streaming Model (GSM) \citep{Peebles,Fisher95,KoppUA,Hamaustheo,Cai16} to model the 2D galaxy-void correlation function in redshift space. 

\medskip
\noindent 
The analyses that first constraints the growth rate around voids from galaxy surveys are \cite{HamausSDSS}, where the authors used the CMASS sample of the Sloan Digital Sky Survey (SDSS), \cite{AB_voids2016}, where we used the low redshift  \textit{6-degree Field Galaxy Survey} (6dFGS) \cite{Jones2004}, and \cite{AH_Viper2016}, where the authors used the high redshift VIPERS survey datasets. While these analyses have shown an overall consistency with the $\Lambda$CDM expectation of the linear growth rate, the GSM does assume a knowledge of the real space density profiles around voids, which may induce a bias in the analysis\footnote{although one can marginalized over the void profiles fitted parameters as it was done in \cite{HamausSDSS}}. In \cite{HamausSDSS2} the authors took advantage of the approximated linear behavior of cosmic void evolution to perform a multipole analysis of the RSD around voids using both the CMASS and the LOWZ galaxy samples of SDSS DR12. Such a mutlipole analysis allows to derive the growth rate purely from the data measurement, assuming a linear relationship between the monopole and the quadrupole (see also the recent work of \cite{Correa} for a complementary approach). With this assumption they have derived a linear growth rate consistent with $\Lambda$CDM in the CMASS sample, but at a $ \sim 2$--$3\sigma$ deviation from it in the LOWZ sample.

\medskip
\noindent In this work we perform an independent analysis from \cite{HamausSDSS2} using a different void finder and a different treatment of the errors which enter into the likelihood analysis. Using the 500 mocks from the  publicly available mock galaxy catalogues produced with the Quick Particle Mesh (QPM) method \cite{QPM}, we test the validity of the mutlipole decomposition and use them to compute the covariance matrix that enters into the likelihood. We will show that in our case, we observe no deviation from $\Lambda$CDM when we disregard the mutlipole measurements at small scales ($<10\;h^{-1}$Mpc). 

\medskip
\noindent This paper is organized as follows: in section \ref{datam} we describe
the data and the mocks we use to perform our analysis, in section \ref{secvoids} we explain how we obtain our void catalogues, in section \ref{sec2} we introduce the model we use to derive the linear growth rate, in sections \ref{secanalyse}, \ref{secres} we test our approach using the QPM mock catalogues and in the CMASS and LOWZ dataset.  In section \ref{conclu} we present our conclusion. 

\section{Data \& Mock Catalogues}\label{datam}
We use the publicly available data of SDSS-III \cite{SDSS3} Data Release 12 (DR12) which contains two datasets of galaxy catalogues from the Baryon Oscillation Spectroscopic Survey (BOSS) \footnote{\url{http://www.sdss3.org/science/boss_publications.php}}: the LOWZ and the CMASS samples. Both map the southern and the northern hemispheres. The LOWZ north/south sample contains $\sim 248/114 \times 10^3$ galaxies in redshift range $0.15<z<0.43$ and a median $\bar{z}=0.32$ while the CMASS north/south sample contains $\sim 569/208 \times 10^3$ galaxies in redshift range $0.43<z<0.70$ with $\bar{z}=0.54$. 

\medskip
\noindent To identify the voids in the galaxy samples and to compute the multipoles, we use the two random catalogues generated by the BOSS collaboration (for each sample e.g. LOWZ north/south and CMASS north/south), featuring the redshift distribution. Each of these catalogues are referred to as RAN and RAN2. These random catalogues are also publicly available and contain about 50 times more points than the observed galaxies. 
\begin{figure}
\includegraphics[width=\columnwidth]{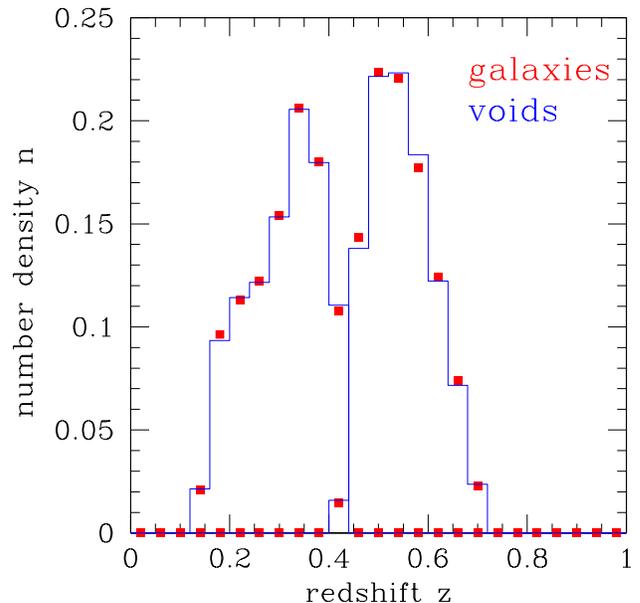} 
\caption{Number density of voids (blue histogram) and galaxies (red squares) in the LOWZ/CMASS samples, as a function of redshift. Both overlap with one another as expected from the void finder.}\label{Figdndz}
\end{figure}

\medskip
\noindent To compute the covariance matrix and to test our analysis, we use the publicly available SDSS-III  DR12  mocks, generated with the Quick Particle Mesh (QPM) algorithm \cite{QPM}. They were generated assuming a flat $\Lambda$CDM cosmology $(\Omega_{\Lambda}=0.71,\Omega_m=0.29, \Omega_b=0.0458,\sigma_8=0.80,h=0.7,n_s=0.97)$ \cite{Chuang}. We note that in these mocks, the linear galaxy bias is $b=2.2$ while in both CMASS and LOWZ it was estimated to be $b=1.85$ \cite{Chuang}.

\section{Void catalogue}\label{secvoids}
To identify the cosmic voids in both the galaxy dataset and the QPM mocks, we use the void finder developed by \cite{Achitouv16}, that was used in the 6dF Galaxy Survey analysis \cite{AB_voids2016} to infer the growth rate. This void finder uses a sample of RAN2 to identify candidate voids with an effective radius $r_v$ that satisfies the following density constrains: $\delta(r_0)<-0.9$; $\delta(r_0+dr)<-0.8$; $\delta(r_0+2dr)<-0.3$ ; $\delta(r_v)>0.15$; $\delta(r_v+dr)>\delta(r_v)$; $\delta(r_v+dr)<0.4$. where the binning is given in steps of $dr=3~h^{-1}$Mpc, $r_0=1.5~h^{-1}$Mpc and $\delta(r)$ is approximated by counting the number of galaxies around each random position we select from RAN2, divided by the number of randoms we compute from the RAN catalogues. The first 3 conditions ensure that the centre of the voids is underdense while the conditions around $r=r_v$ ensure that the selected voids have a ridge. We then perform two loops over these void candidates that satisfy the density conditions: the first loop to smooth the individual void profiles by requiring that $\delta(r+3dr<R<r_v/2)<-0.3$, The second to remove overlapping voids, keeping the largest. 

\medskip
\noindent We use between 5 to 8 times the number of candidate positions as tracers, which is a good compromise between numerical computing power and having a convergence in the number of identified voids. Indeed, given that we remove overlapping voids, increasing the number of candidates can increase the number of identified voids up to a limited number. Keeping the same criteria for the data samples and the mocks, we end up with a selection of voids distributed in redshift as displayed in Fig.
\ref{Figdndz}. 

\begin{figure}
\includegraphics[width=\columnwidth]{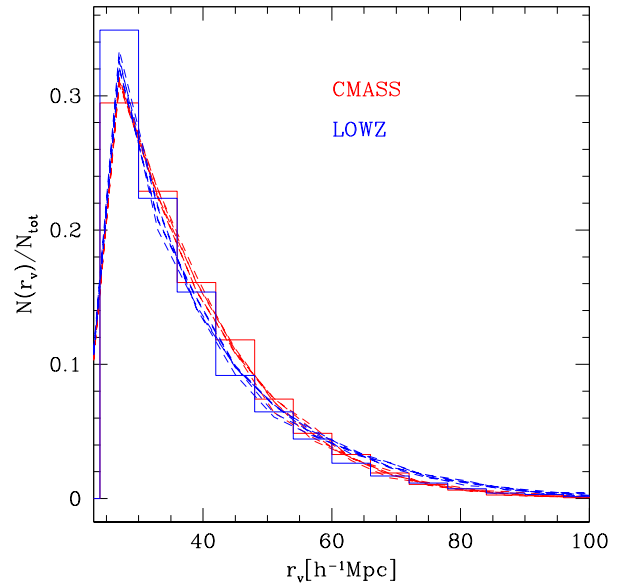} 
\caption{Normalized number of voids as a function of the void radius in the LOWZ/CMASS samples (histograms) and in 5 randomly selected mocks (dashed curves). These PDFs are qualitatively similar as a result of the void finder criteria, which remain unchanged in both samples.}\label{FigdndR}
\end{figure}

\medskip
\noindent We repeat the same procedure using 500 QPM mocks for LOWZ North/South and CMASS North/South. Finally, we introduce a cut in the minimum size of the voids for the RSD analysis $r_{v}^{\rm min}=25~h^{-1}$Mpc. The motivation for this cut is that $(i)$ small voids identified with galaxy tracers do not necessarily correspond to underdensities in the  matter density field. They also show a stronger deviation from linear evolution, and the galaxy bias around small voids can be amplified compared to the large-scale average bias  \citep{Sutter13,Pollina16,PollinaDES}.  $(ii)$ we found that the overall void size distribution matches the mean value of the QPM mocks distribution when $r>r_{v}^{\rm min}$. Although we are not interested in testing for void abundance in this work, having a mismatch in the void size distribution could introduce an offset between the mean void density profiles measured in the data and in the mocks, which could possibly introduce a bias in the derivation of our cosmological parameters. After applying this threshold, we found a total of $5986$ voids identified in the LOWZ sample and $6373$ in the CMASS sample. The normalized number of voids as a function of radius is displayed in Fig.\ref{FigdndR}. The blue/red histograms correspond to the LOWZ/CMASS samples while the blue/red dashed curves correspond to 5 randomly selected samples from LOWZ/CMASS mocks, respectively. The mean void radius in the LOWZ/CMASS samples are, respectively, $r_v=38.5$ and $38~h^{-1}$Mpc. 

\section{Methodology}\label{sec2}
\subsection{Multipole decomposition}
The peculiar velocities of galaxies, $\mathbf{v}$, that are due to the local
gravitational potential result, on small scales, in random motions of
galaxies within virialized halos. In principle this effect is not present within voids, which are generally empty of galaxies in their centre. On large scales however, the coherent bulk flow pointing outwards from centres of voids is responsible for an overall coherent
distortion known as the `Kaiser effect' \cite{Kaiser1987}. It is this coherent outflow that carries information of the linear growth rate. Indeed, the galaxy peculiar velocities sourced by the underlying mass distribution of a void can be expressed in the linear regime as (\citep{Peebles,Hamaustheo,Cai16,HamausSDSS2}):

\begin{equation}
\mathbf{v}(\mathbf{r})=\frac{-1}{3}\frac{f(z)H(z)}{1+z} \mathbf{r}\Delta(r)
\end{equation}
where $f(z)$ is the linear growth rate, $H(z)$ is the Hubble rate, $\mathbf{r}\equiv \mathbf{x}-\mathbf{X}$ is the separation between the comoving coordinate of the void centre $\mathbf{X} $, and a galaxy at position $\mathbf{x}$. We also assume that on average the void density profiles are spherical and can be described by the density contrast $\Delta(r)$ where $r\equiv |~\mathbf{r}~|$. To relate the averaged galaxy density contract, $\bar{\xi}(r)$, to the
matter density contrast, we generally assume a linear bias $b$ such that $\bar{\xi}(r)=b\Delta(r)$ and
\begin{equation}
\bar{\xi}(r)\equiv \frac{3}{r^3}\int_{0}^{r}\xi(y)y^2dy
\end{equation}
where $\xi(r)$ is equivalent to the galaxy density contrast at a scale $r$ (i.e. the galaxy-void cross-correlation function). 

\medskip
\noindent The peculiar velocity of a galaxy gives a contribution to the redshift space separation between the galaxy and the void centre, and in the limit where $\mid \mathbf{r}\mid \ll \mathbf{X}$, 
\begin{equation}
\mathbf{s} = \mathbf{r} + \frac{(1+z) \mathbf{\hat{X}.v} }{H(z)}
\mathbf{\hat{X}} ,
\label{RSDeq}
\end{equation}
where $\mathbf{\hat{X}}$ is the unitary vector along the line of sight to our void centre.

\noindent Performing a Jacobian transformation between the coordinate $\mathbf{s}$ and $\mathbf{r}$, at linear order, the redshift-space 2D correlation function can be described by
(\cite{Kaiser1987,Hamilton1992,Cai16,HamausSDSS2}):
\begin{equation}
\xi^s(r,\mu) = \xi_0(r)+\frac{3\mu^2-1}{2}\xi_2(r)
\label{poly}
\end{equation} 
where $\mu \equiv \cos(\theta)=\hat{\mathbf{X}}.\hat{\mathbf{r}}$ is the cosine of the angle between the line-of-sight direction and the separation vector while $\xi_0,\xi_2$ are the monopole and the quadrupole respectively, computed using the Legendre polynomials $P_l(\mu)$ via  
\begin{equation}
\xi_{l}(r)=\int_{0}^{1}\xi^{s}(r,\mu)(1+2l)P_{l}(\mu)d\mu.\label{decomp}
\end{equation}

\noindent  In the linear regime \cite{Kaiser1987},
\begin{align*}
\xi_0(r) & = \left(1 + \frac{\beta}{3}\right) \xi(r) \\
\xi_2(r) & = \frac{2\beta}{3}\left( \xi(r) - \bar{\xi}(r) \right) ,
\end{align*}
where $\beta = f/b$ and $\xi(r)$ is the real-space galaxy-void correlation function. These expressions lead to a simple relationship between monopole and quadrupole:
\begin{equation}
\xi_0(r)-\bar{\xi_0}(r)-\xi_2(r)\frac{3+\beta}{2\beta}=0\label{eupsi}
\end{equation}

\noindent This is the key equation that \cite{HamausSDSS2} have used to probe $\beta$ solely by measuring the monopole and the quadrupole. We will also use this equation in what follows, but we will introduce a cut at the scale $r_{\rm cut}$ below which this approximation is no longer valid. 

\subsection{Measurement of the galaxy-void correlation function }
To perform the mutlipole decomposition we start by measuring the void-tracer cross-correlation functions using the Landy-Szalay estimator:
\begin{equation}
\xi_{vg}^{s}(r,\mu) = \frac{N_{rg}N_{rv}}{R_vR_g} \left( \frac{D_vD_g}{N_gN_v}
- \frac{D_gR_v}{N_gN_{rv}} - \frac{D_vR_g}{N_vN_{rg}} \right) + 1 ,
\label{LSeq}
\end{equation}
where $D_vD_g$ is the number of data void-galaxy pairs, $R_v R_g$ the
random void-galaxy pairs and $D_{g/v}R_{g/v}$ the number of
galaxy/void data-random pairs, in bins at separation $r$ and $\mu$.  The total
number of galaxies, voids, galaxy-randoms and void-randoms are $N_g$,
$N_v$, $N_{rg}$ and $N_{rv}$, respectively. In all cases we use a sample of the first random catalogues provided by the BOSS collaboration, having 10 times the number of galaxies/voids than our data samples. 

\medskip
\noindent
We measure Eq.~\ref{LSeq} in bins of $d\mu=0.045$ and $dr=4~h^{-1}$Mpc.

\subsection{The likelihood analysis}
\begin{figure}
\begin{tabular}{cc}
\includegraphics[width=\columnwidth]{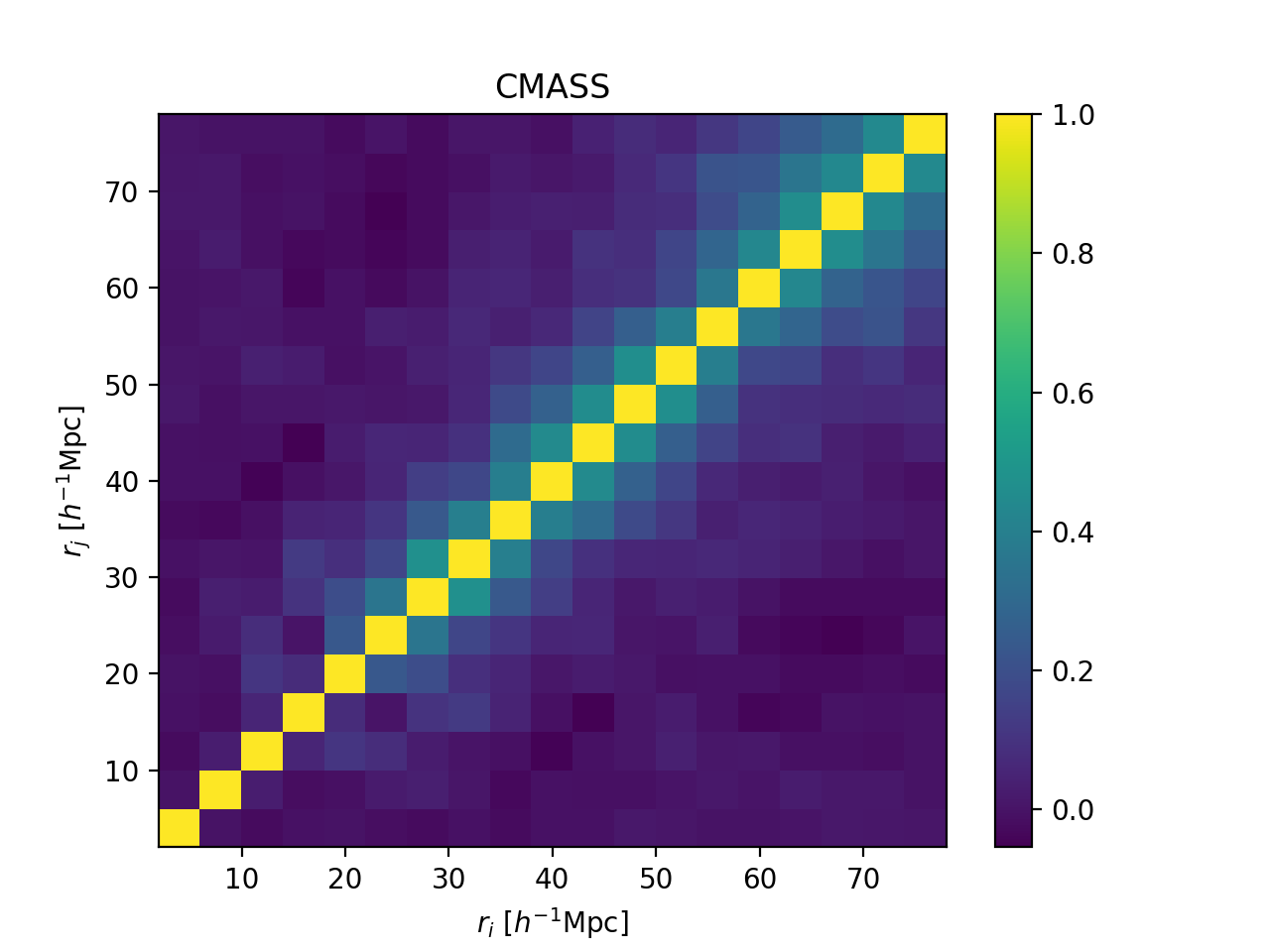}\\
\includegraphics[width=\columnwidth]{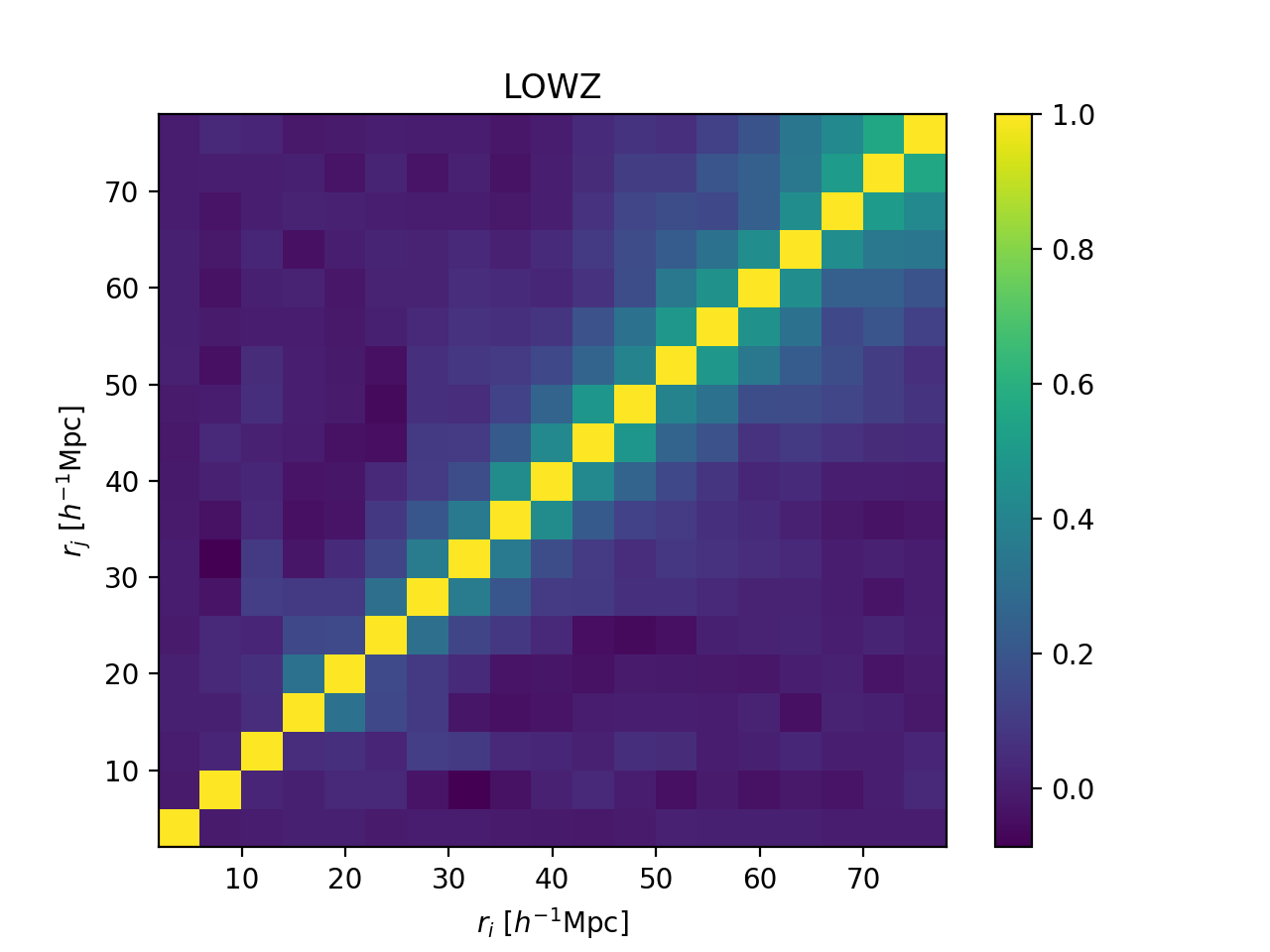}
\end{tabular}
\caption{Correlation matrix for the residuals (left hand side of Eq.~\ref{eupsi}), in the CMASS (top) and LOWZ (bottom) mocks.}\label{Figcov} 
\end{figure}

To infer the linear growth rate from the measurement of the monopole and quadrupole, we solve for the value of $\beta$ which satisfies Eq.~\ref{eupsi}, performing a Gaussian likelihood

\begin{equation}
L(\xi_0,\xi_2\mid\beta)=\frac{1}{(2\pi)^{N/2}\sqrt{\rm det \mathbf{C}}}\exp{\left[-\frac{1}{2}\sum_{i,j=l_{\rm min}}^{N}   \varepsilon_i \mathbf{C}_{ij}^{-1}\varepsilon_j\right]}, 
\label{Eqlike}
\end{equation}
where the sum is in radial bins $r_i=\left[r_{\rm cut}, r_{\rm max}\right]$, $\varepsilon_i \equiv \xi_0(r_i)-\bar{\xi_0}(r_i)-\xi_2(r_i)\frac{3+\beta}{2\beta}$ is the left hand side of Eq.~\ref{eupsi} and $\mathbf{C}$ is the covariance matrix $\mathbf{C}_{ij}=\left<\varepsilon_i\varepsilon_j\right>$ which depends explicitly on $\beta$. Hence the normalization of the likelihood needs to be taken into account. Unlike the analysis performed in \cite{HamausSDSS2}, which uses a jackknife method to estimate the covariance matrix, in what follows we compute the covariance matrix using 500 QPM mocks. 

\medskip
\noindent In Fig.~\ref{Figcov} we show the correlation matrix (covariance matrix of the residuals after normalization by its diagonal) which can be compared to Fig.~4 in \cite{HamausSDSS2}. The correlations between our bins follow the same qualitative trend as \cite{HamausSDSS2}: in the inner part of the voids, $r/r_v<1$, the bins seem less correlated while for $r/r_v>1$ we see some off diagonal correlations. 
We also note that in order to compute the galaxy-void correlation function we employ the Landy-Szalay (LS) estimator, while \cite{HamausSDSS2} use the approximation $\xi_l(r)\simeq \left<D_vD_g\right>-\left<D_vR_g\right>$.

\section{Analysis}\label{secanalyse}
\begin{figure}
\begin{tabular}{cc}
\includegraphics[width=\columnwidth]{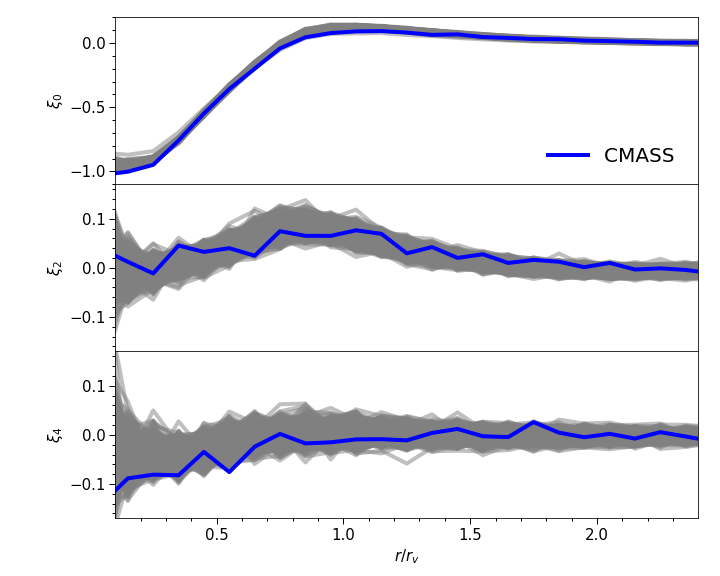}\\
\includegraphics[width=\columnwidth]{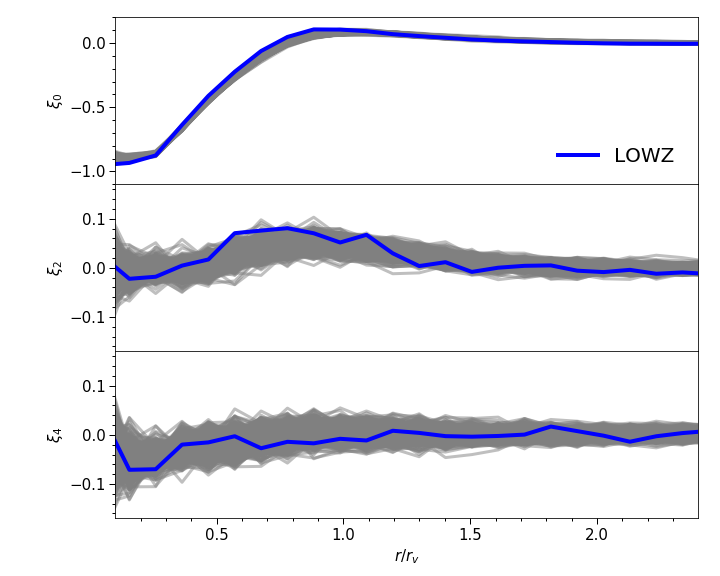} 
\end{tabular}
\caption{Multipole measurements in the mocks (grey curves) and in the data (blue curves) we obtained from Eq.~\ref{decomp}. The data multipoles are qualitatively in good agreement with the ones we obtain in the mocks.}\label{Figmulti}
\end{figure}
\begin{figure}
\begin{tabular}{cc}
\includegraphics[width=\columnwidth]{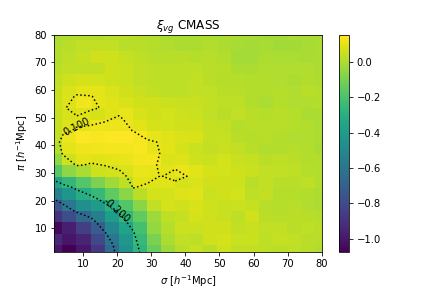}\\
\includegraphics[width=\columnwidth]{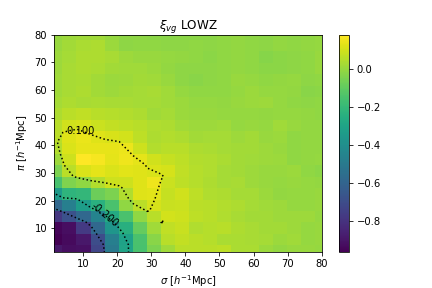} 
\end{tabular}
\caption{The mean measurement of the 2D void-galaxy correlation
  function in CMASS (upper panel) and in LOWZ sample
  (lower panel). The dotted
  lines show iso-contours of the data. Both measurements show apparent asymetries that are characteristics of the coherent outflow of galaxies, sourced by the gravitational potential of the voids.}\label{Figxi2D}
\end{figure}

We start by using Eq. \ref{LSeq} to measure the galaxy-void correlation function in the data and the mocks, and then we apply Eq. \ref{decomp} to compute the monopole ($l$=0), quadrupole ($l$=2) and hexadecapole ($l$=4).The resulting multipoles are shown in Fig.~\ref{Figmulti}, where the grey curves correspond to the mocks measurements (1000 in total for CMASS and LOWZ) and the blue curves to the data. First we observe that the multipoles computed from the data and the mocks are qualitatively in good agreement with one another. Second we observe that for $r/r_v\leq 0.3 $, which corresponds to a radius below $r\sim 10\;h^{-1}$Mpc, the slope of the monopole changes, and $\xi_0 \rightarrow -1$ while $| \xi_4|>0$. These behaviours could indicate a breakdown of the linear assumptions and/or ill-defined regions due to the lack of particle counts at the core of the voids. In any case, these low scales can not be used within our current linear model hypothesis. Hence in what follows we define a cut-off scale {$r_{\rm cut}$} below which we disregard our measurements when performing the likelihood analysis.

\medskip
Finally, we also show the measurement of the 2D galaxy-void correlation function in both LOWZ/CMASS samples in Fig.~\ref{Figxi2D}, that we have measured parallel ($\pi$) and perpendicular ($\sigma$) to the the line of sight, using a binning of $4\;h^{-1}$Mpc. This is just to illustrate the asymmetry due to the peculiar velocities of galaxies that have a coherent outflow due to the gravitational potential of the void. This measurement could be used to extract the growth rate using a quasi-linear modelling (e.g. Gaussian Streaming Model), as it was done in \citep{HamausSDSS,AB_voids2016,AH_Viper2016}. However it would require assumptions on the real space density profiles around the voids, which we know are sensitive to the underlying cosmology and to the void finder algorithm. Hence we do not explore further these 2D measurements.

\section{Results}\label{secres}
In what follows, we set $r_{\rm cut}=10\;h^{-1}$Mpc and we use our measurement in bins of $dr=4\;h^{-1}$Mpc up to $r_{\rm max}=78\;h^{-1}$Mpc. We have verified that the results we present in this section remain unchanged via the transformation $r_{\rm cut}\rightarrow r_{\rm cut}\pm dr$ or $r_{\rm max}\rightarrow r_{\rm max}\pm dr$. We also tested that the inferred value of $\beta$ is insensitive to the fiducial size of our voids $r_v$, nor to the hemisphere (splitting voids in large vs. small, separating north vs. south datasets). Thus for this analysis, we combined all the void sizes to obtain better statistical errors. 

\noindent To obtain the best fit value for $\beta$, we use a large prior of $\beta=\left[-0.1,1.2\right]$ in steps of $d\beta=0.0024$. We have verified that our results remain unchanged by increasing the prior range. 
\subsection{Mocks}\label{secm}
\begin{figure}
\includegraphics[width=\columnwidth]{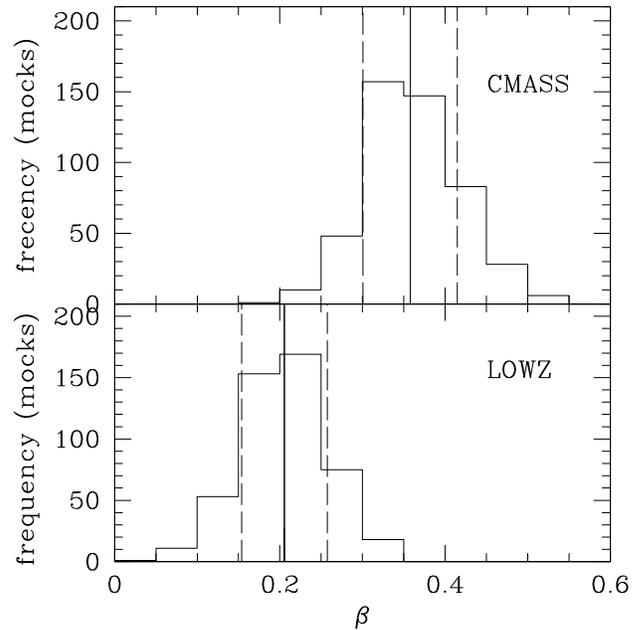} 
\caption{Histogram of the inferred values of $\beta$ using the galaxy-void multipole analysis in the QPM mocks. The solid lines correspond to the mean value $\bar{\beta}=0.36,0.21$  and the dashed lines to the 1-$\sigma$ deviation $0.06,0.05$ for the CMASS, LOWZ samples, respectively.}\label{FigBFmocks}
\end{figure}
We start our analysis by inferring the value of $\beta$ on each individual mock catalogue, using the Likelihood computation given in Eq.~\ref{Eqlike} in order to evaluate the uncertainties on the $\beta$ measurement. In Fig.~\ref{FigBFmocks} we show the histogram of the best fit values we have found in the CMASS, LOWZ mocks as well as the mean values and the standard deviation: $\bar{\beta}=0.36\pm 0.06$ for the CMASS mocks and $\bar{\beta}=0.21\pm0.05$ for the LOWZ mocks. It is not trivial to compare these values to the mock expectations. Indeed, given the fiducial cosmology of the QPM mocks \cite{QPM} we can easily compute the expected value for the growth rate but the linear bias is not explicitly given at the mean redshift of the mocks. At $z=0.5$ the linear bias is expected to be $b=2.2$ for the QPM mocks \cite{SDSS3}. In such case, we can extrapolate the value $\beta(z=0.5)=0.34$. This value can be compared to the CMASS mocks because in these mocks the redshift is $z\sim 0.54$. If we neglect the redshift dependence of the linear bias and keep $b=2.2$ but use the growth rate at the redshift of the LOWZ mocks then we can expect a value of $\beta=0.30$.  Both theoretical values are within 2-$\sigma$ deviation from the mean of $\beta$ we obtain. Finally before performing the data analysis, we should make a few critical remarks:
\begin{itemize}
\item Galaxies around voids may be more biased compared to the average galaxies in the full simulation. In which case we can expect the fiducial value of $\beta$ to be lower than the one computed from $b=2.2$. We note however that in \cite{AB_voids2016} the value of the linear bias we have inferred in mocks using the galaxy-void and the galaxy-galaxy correlation functions were consistent with one another. This must depend on the fiducial void size and the characteristics of the void profiles (e.g amplitude at the void ridge).

\item We note that if we would have inferred $\beta$ from the mocks mean measurement of the multipole, the systematic errors due to the linear assumptions would most likely dominate: in the LOWZ sample we have found the mean of the $\beta$ best fit values to be in agreement with the fiducial cosmology at 2-, but not 1-$\sigma$. This may be an issue for upcoming surveys such as TAIPAN which will probe a larger volume, with a higher density of galaxies and voids \cite{TaipanWP} at low redshift.

\item We also point out the limitation of using QPM mocks \cite{QPM} to test for the validity of the growth rate at low redshift. Indeed, unlike in full N-body simulations, efficient algorithms such as \cite{QPM} have not yet fully investigated the validity of their approach to reproduce the statistical description of the undersense matter density field.

\end{itemize}
 
\noindent Overall, apart from these remarks, we find that the mean of $\beta$ from the best fit values of the mocks are within 2-$\sigma$ deviation of the expected fiducial cosmology, which validate our approach given the statistical errors we have.

\subsection{Data}
\begin{figure}
\includegraphics[width=\columnwidth]{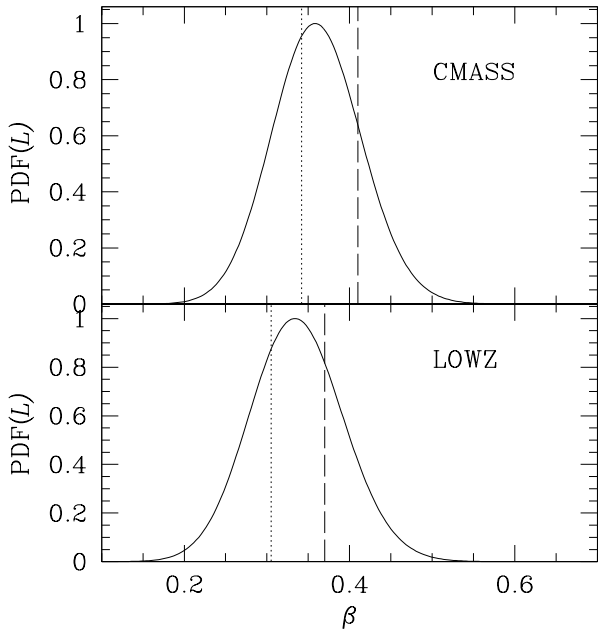} 
\caption{Posterior distribution for $\beta$ in the CMASS and LOWZ data. The long-dashed line corresponds to the expected $\Lambda$CDM cosmology with a linear bias $b=1.85$ as it was inferred in \cite{Chuang} while the dotted line corresponds to the $\Lambda$CDM cosmology with $b=2.2$ (i.e., the value of the linear bias in the QPM mocks at $z=0.5$). }\label{FigBFdata}
\end{figure}

Following the same procedure but for the data sample, we find a best fit $\beta=0.33\pm0.06$ and $\beta=0.36\pm0.05$ for the LOWZ and CMASS samples, respectively, with a reduced $\chi^2/d.o.f.$ of $22.6/16=1.41$ and $21.8/16=1.36$. The posterior distribution is shown in Fig.~\ref{FigBFdata}. We note that our errors on $\beta$ are consistent with what we found using the standard deviation of the best fit values from the mocks and that the best fit values correspond to the mean value of the likelihood PDF.

\medskip
\noindent Once again we can compare these results with the expected values of $\beta$ in the case of a $\Lambda$CDM cosmology (see sec. \ref{datam} for cosmological parameter values). With a linear bias $b=1.85$ (as inferred in \cite{Chuang}), the theoretical values for LOWZ/CMASS are $\beta=0.37,0.41$ respectively. These are the same reference values that \cite{HamausSDSS2} have used to compare with their results. In Fig.~\ref{FigBFdata} they correspond to the long dashed lines. Unlike what the authors in \cite{HamausSDSS2} have found, we obtain a 1-$\sigma$ agreement with respect to $\Lambda$CDM, both for the LOWZ and the CMASS samples. We also show in Fig.~\ref{FigBFdata} the fiducial values of $\beta$ for $b=2.2$ (motivated by the discussion in sec.~\ref{secm}). The latter is also consistent at 1-$\sigma$ with our best fitting values. 

\section{Conclusion}\label{conclu}
In this work we have probed the parameter $\beta=f/b$, using the public galaxy catalogues released by the BOSS collaboration and an RSD mutlipole analysis of the galaxy-void cross-correlation function. The model we used to infer the growth rate is derived from linear theory and was initially used in \cite{HamausSDSS2} to perform a similar analysis. However, in this work we find that our derived values for the growth rate are consistent with a $\Lambda$CDM cosmology within 1-$\sigma$. The main differences in this analysis compared to the one presented in \cite{HamausSDSS2} are:

\begin{itemize}
\item Our void catalogues are completely independent and based on different criteria (density criteria \cite{Achitouv16} vs. watershed transform \cite{VIDE}). While the peak of the void size distribution is relatively similar in both studies, we have better statistics on the number of voids in the LOWZ sample. As a result, our errors on $\beta$ are similar in both the LOWZ $\Delta \beta = 0.06$ and CMASS sample $\Delta \beta = 0.05$.

\item Motivated by our analysis with the mocks, we introduce a cut in scale to disregard our measurement at the centre of the voids where $|\delta|\rightarrow -1$, which corresponds to the non-linear regime where Eq.~\ref{eupsi} does not hold in principle, as we discuss in sec.~\ref{secanalyse}.

\item The treatment of the covariance matrix is different: in this work we used the mocks to compute the covariance while in \cite{HamausSDSS2} they used a jackknife method. We also provide in this work a complete study of the inferred values of $\beta$ within the mocks in order to check the validity of our model (sec.~\ref{secm}). 

\end{itemize}

\medskip
\noindent Overall, this work has provided some interesting results: 
\begin{itemize}
\item Using mock catalogues, we have shown that $\beta$ can be extracted using no theoretical modelling of the void-galaxy correlation function in real space. This is particularly interesting to avoid assuming a fiducial cosmology in order to predict the void density profiles, which could lead to potential bias of the growth rate value (the void density profiles carry the imprints of the cosmology e.g. \cite{ABPW,Adermann,Kreisch,Paillas,Cautun}), or to avoid parametrizing the real space density profile and/or marginalising over the profile parameters, which would introduce potentially weaker constraints on the growth rate.

\item The values of $\beta$ that we obtain in the LOWZ/CMASS datasets are consistent with the value probed in \cite{Chuang}. However in \cite{Chuang} the scale range used to derive $\beta$ is $\left[40-180\right]~h^{-1}$Mpc, while we used the information contained within ranges $\left[10-78\right]~h^{-1}$Mpc. This illustrates again the complementarity of using cosmic voids to perform cosmological analysis: we have access to additional information, and the systematic errors are different. 
\end{itemize}

\noindent Finally we can emphasise on the fact that that the value of $\beta$ we obtained in this analysis is in good agreement with the $\Lambda$CDM linear prediction. It would be interesting to probe the information contained in smaller scales (e.g. below $10~h^{-1}$Mpc) where the non-linearities can carry more information. For instance, in \cite{AchitouvCai} we have shown how the growth rate of cosmic structure can vary considerably when the underlying matter density $|\Delta|\geq 1$. We hope to perform such analysis in future work.

\section*{Acknowledgments}
I am very grateful to Nico Hamaus and Cai Yanchuan for their useful discussions and for carefully proofreading this manuscript,  giving me pertinent feedback.  
The research leading to these results has received funding from the European Research Council under the European Community Seventh Framework Programme (FP7/2007-2013 Grant Agreement no.  279954) RC-StG \textit{EDECS}.


\bibliographystyle{chicago}  
\bibliography{IA_SDSSRSDv3.bib}

\begin{thebibliography}{}

\bibitem[\protect\citeauthoryear{{Achitouv}}{{Achitouv}}{2016}]{Achitouv16}
{Achitouv}, I. (2016, November).
\newblock {Testing the imprint of nonstandard cosmologies on void profiles
  using Monte Carlo random walks}.
\newblock {\em prd\/}~{\em 94\/}(10), 103524.

\bibitem[\protect\citeauthoryear{{Achitouv}}{{Achitouv}}{2017}]{Achitouv17}
{Achitouv}, I. (2017, October).
\newblock {Improved model of redshift-space distortions around voids:
  Application to quintessence dark energy}.
\newblock {\em prd\/}~{\em 96\/}(8), 083506.

\bibitem[\protect\citeauthoryear{{Achitouv}, {Baldi}, {Puchwein}, and
  {Weller}}{{Achitouv} et~al.}{2016}]{ABPW}
{Achitouv}, I., M.~{Baldi}, E.~{Puchwein}, and J.~{Weller} (2016, May).
\newblock {Imprint of f (R ) gravity on nonlinear structure formation}.
\newblock {\em prd\/}~{\em 93\/}(10), 103522.

\bibitem[\protect\citeauthoryear{{Achitouv}, {Blake}, {Carter}, {Koda}, and
  {Beutler}}{{Achitouv} et~al.}{2017}]{AB_voids2016}
{Achitouv}, I., C.~{Blake}, P.~{Carter}, J.~{Koda}, and F.~{Beutler} (2017,
  April).
\newblock {Consistency of the growth rate in different environments with the
  6-degree Field Galaxy Survey: Measurement of the void-galaxy and
  galaxy-galaxy correlation functions}.
\newblock {\em prd\/}~{\em 95\/}(8), 083502.

\bibitem[\protect\citeauthoryear{{Achitouv} and {Cai}}{{Achitouv} and
  {Cai}}{2018}]{AchitouvCai}
{Achitouv}, I. and Y.-C. {Cai} (2018, November).
\newblock {Modeling the environmental dependence of the growth rate of cosmic
  structure}.
\newblock {\em \prd\/}~{\em 98}, 103502.

\bibitem[\protect\citeauthoryear{{Achitouv}, {Neyrinck}, and
  {Paranjape}}{{Achitouv} et~al.}{2015}]{ANP}
{Achitouv}, I., M.~{Neyrinck}, and A.~{Paranjape} (2015, August).
\newblock {Testing spherical evolution for modelling void abundances}.
\newblock {\em mnras\/}~{\em 451}, 3964--3974.

\bibitem[\protect\citeauthoryear{{Adermann}, {Elahi}, {Lewis}, and
  {Power}}{{Adermann} et~al.}{2018}]{Adermann}
{Adermann}, E., P.~J. {Elahi}, G.~F. {Lewis}, and C.~{Power} (2018, October).
\newblock {Cosmic voids in evolving dark sector cosmologies: the high-redshift
  universe}.
\newblock {\em mnras\/}~{\em 479}, 4861--4877.

\bibitem[\protect\citeauthoryear{{Alam}, {Ata}, {Bailey}, {Beutler}, {Bizyaev},
  {Blazek}, {Bolton}, {Brownstein}, {Burden}, {Chuang}, {Comparat}, {Cuesta},
  {Dawson}, {Eisenstein}, {Escoffier}, {Gil-Mar{\'\i}n}, {Grieb}, {Hand}, {Ho},
  {Kinemuchi}, {Kirkby}, {Kitaura}, {Malanushenko}, {Malanushenko}, {Maraston},
  {McBride}, {Nichol}, {Olmstead}, {Oravetz}, {Padmanabhan},
  {Palanque-Delabrouille}, {Pan}, {Pellejero-Ibanez}, {Percival}, {Petitjean},
  {Prada}, {Price-Whelan}, {Reid}, {Rodr{\'\i}guez-Torres}, {Roe}, {Ross},
  {Ross}, {Rossi}, {Rubi{\~n}o-Mart{\'\i}n}, {Saito}, {Salazar-Albornoz},
  {Samushia}, {S{\'a}nchez}, {Satpathy}, {Schlegel}, {Schneider},
  {Sc{\'o}ccola}, {Seo}, {Sheldon}, {Simmons}, {Slosar}, {Strauss}, {Swanson},
  {Thomas}, {Tinker}, {Tojeiro}, {Maga{\~n}a}, {Vazquez}, {Verde}, {Wake},
  {Wang}, {Weinberg}, {White}, {Wood-Vasey}, {Y{\`e}che}, {Zehavi}, {Zhai}, and
  {Zhao}}{{Alam} et~al.}{2017}]{SDSS3}
{Alam}, S., M.~{Ata}, S.~{Bailey}, F.~{Beutler}, D.~{Bizyaev}, J.~A. {Blazek},
  A.~S. {Bolton}, J.~R. {Brownstein}, A.~{Burden}, C.-H. {Chuang},
  J.~{Comparat}, A.~J. {Cuesta}, K.~S. {Dawson}, D.~J. {Eisenstein},
  S.~{Escoffier}, H.~{Gil-Mar{\'\i}n}, J.~N. {Grieb}, N.~{Hand}, S.~{Ho},
  K.~{Kinemuchi}, D.~{Kirkby}, F.~{Kitaura}, E.~{Malanushenko},
  V.~{Malanushenko}, C.~{Maraston}, C.~K. {McBride}, R.~C. {Nichol}, M.~D.
  {Olmstead}, D.~{Oravetz}, N.~{Padmanabhan}, N.~{Palanque-Delabrouille},
  K.~{Pan}, M.~{Pellejero-Ibanez}, W.~J. {Percival}, P.~{Petitjean},
  F.~{Prada}, A.~M. {Price-Whelan}, B.~A. {Reid}, S.~A.
  {Rodr{\'\i}guez-Torres}, N.~A. {Roe}, A.~J. {Ross}, N.~P. {Ross}, G.~{Rossi},
  J.~A. {Rubi{\~n}o-Mart{\'\i}n}, S.~{Saito}, S.~{Salazar-Albornoz},
  L.~{Samushia}, A.~G. {S{\'a}nchez}, S.~{Satpathy}, D.~J. {Schlegel}, D.~P.
  {Schneider}, C.~G. {Sc{\'o}ccola}, H.-J. {Seo}, E.~S. {Sheldon},
  A.~{Simmons}, A.~{Slosar}, M.~A. {Strauss}, M.~E.~C. {Swanson}, D.~{Thomas},
  J.~L. {Tinker}, R.~{Tojeiro}, M.~V. {Maga{\~n}a}, J.~A. {Vazquez},
  L.~{Verde}, D.~A. {Wake}, Y.~{Wang}, D.~H. {Weinberg}, M.~{White}, W.~M.
  {Wood-Vasey}, C.~{Y{\`e}che}, I.~{Zehavi}, Z.~{Zhai}, and G.-B. {Zhao} (2017,
  Sep).
\newblock {The clustering of galaxies in the completed SDSS-III Baryon
  Oscillation Spectroscopic Survey: cosmological analysis of the DR12 galaxy
  sample}.
\newblock {\em mnras\/}~{\em 470}, 2617--2652.

\bibitem[\protect\citeauthoryear{{Beutler}, {Blake}, {Colless}, {Jones},
  {Staveley-Smith}, {Poole}, {Campbell}, {Parker}, {Saunders}, and
  {Watson}}{{Beutler} et~al.}{2012}]{Beutler6dF}
{Beutler}, F., C.~{Blake}, M.~{Colless}, D.~H. {Jones}, L.~{Staveley-Smith},
  G.~B. {Poole}, L.~{Campbell}, Q.~{Parker}, W.~{Saunders}, and F.~{Watson}
  (2012, July).
\newblock {The 6dF Galaxy Survey: z{\ap} 0 measurements of the growth rate and
  {$\sigma$}$_{8}$}.
\newblock {\em mnras\/}~{\em 423}, 3430--3444.

\bibitem[\protect\citeauthoryear{{Blake}, {Brough}, {Colless}, {Contreras},
  {Couch}, {Croom}, {Croton}, {Davis}, {Drinkwater}, {Forster}, {Gilbank},
  {Gladders}, {Glazebrook}, {Jelliffe}, {Jurek}, {Li}, {Madore}, {Martin},
  {Pimbblet}, {Poole}, {Pracy}, {Sharp}, {Wisnioski}, {Woods}, {Wyder}, and
  {Yee}}{{Blake} et~al.}{2012}]{Blake2011}
{Blake}, C., S.~{Brough}, M.~{Colless}, C.~{Contreras}, W.~{Couch}, S.~{Croom},
  D.~{Croton}, T.~M. {Davis}, M.~J. {Drinkwater}, K.~{Forster}, D.~{Gilbank},
  M.~{Gladders}, K.~{Glazebrook}, B.~{Jelliffe}, R.~J. {Jurek}, I.-h. {Li},
  B.~{Madore}, D.~C. {Martin}, K.~{Pimbblet}, G.~B. {Poole}, M.~{Pracy},
  R.~{Sharp}, E.~{Wisnioski}, D.~{Woods}, T.~K. {Wyder}, and H.~K.~C. {Yee}
  (2012, September).
\newblock {The WiggleZ Dark Energy Survey: joint measurements of the expansion
  and growth history at z $lt$ 1}.
\newblock {\em mnras\/}~{\em 425}, 405--414.

\bibitem[\protect\citeauthoryear{{Cai}, {Padilla}, and {Li}}{{Cai}
  et~al.}{2015}]{Cai15}
{Cai}, Y.-C., N.~{Padilla}, and B.~{Li} (2015, Jul).
\newblock {Testing gravity using cosmic voids}.
\newblock {\em mnras\/}~{\em 451}, 1036--1055.

\bibitem[\protect\citeauthoryear{{Cai}, {Taylor}, {Peacock}, and
  {Padilla}}{{Cai} et~al.}{2016}]{Cai16}
{Cai}, Y.-C., A.~{Taylor}, J.~A. {Peacock}, and N.~{Padilla} (2016, November).
\newblock {Redshift-space distortions around voids}.
\newblock {\em mnras\/}~{\em 462}, 2465--2477.

\bibitem[\protect\citeauthoryear{{Cautun}, {Paillas}, {Cai}, {Bose}, {Armijo},
  {Li}, and {Padilla}}{{Cautun} et~al.}{2018}]{Cautun}
{Cautun}, M., E.~{Paillas}, Y.-C. {Cai}, S.~{Bose}, J.~{Armijo}, B.~{Li}, and
  N.~{Padilla} (2018, May).
\newblock {The Santiago-Harvard-Edinburgh-Durham void comparison - I. SHEDding
  light on chameleon gravity tests}.
\newblock {\em mnras\/}~{\em 476}, 3195--3217.

\bibitem[\protect\citeauthoryear{{Chuang}, {Pellejero-Ibanez},
  {Rodr{\'\i}guez-Torres}, {Ross}, {Zhao}, {Wang}, {Cuesta},
  {Rubi{\~n}o-Mart{\'\i}n}, {Prada}, {Alam}, {Beutler}, {Eisenstein},
  {Gil-Mar{\'\i}n}, {Grieb}, {Ho}, {Kitaura}, {Percival}, {Rossi},
  {Salazar-Albornoz}, {Samushia}, {S{\'a}nchez}, {Satpathy}, {Slosar},
  {Thomas}, {Tinker}, {Tojeiro}, {Vargas-Maga{\~n}a}, {Vazquez}, {Brownstein},
  {Nichol}, and {Olmstead}}{{Chuang} et~al.}{2017}]{Chuang}
{Chuang}, C.-H., M.~{Pellejero-Ibanez}, S.~{Rodr{\'\i}guez-Torres}, A.~J.
  {Ross}, G.-b. {Zhao}, Y.~{Wang}, A.~J. {Cuesta}, J.~A.
  {Rubi{\~n}o-Mart{\'\i}n}, F.~{Prada}, S.~{Alam}, F.~{Beutler}, D.~J.
  {Eisenstein}, H.~{Gil-Mar{\'\i}n}, J.~N. {Grieb}, S.~{Ho}, F.-S. {Kitaura},
  W.~J. {Percival}, G.~{Rossi}, S.~{Salazar-Albornoz}, L.~{Samushia}, A.~G.
  {S{\'a}nchez}, S.~{Satpathy}, A.~{Slosar}, D.~{Thomas}, J.~L. {Tinker},
  R.~{Tojeiro}, M.~{Vargas-Maga{\~n}a}, J.~A. {Vazquez}, J.~R. {Brownstein},
  R.~C. {Nichol}, and M.~D. {Olmstead} (2017, Oct).
\newblock {The clustering of galaxies in the completed SDSS-III Baryon
  Oscillation Spectroscopic Survey: single-probe measurements from DR12 galaxy
  clustering - towards an accurate model}.
\newblock {\em mnras\/}~{\em 471}, 2370--2390.

\bibitem[\protect\citeauthoryear{{Clampitt}, {Cai}, and {Li}}{{Clampitt}
  et~al.}{2013}]{Clampitt2012}
{Clampitt}, J., Y.-C. {Cai}, and B.~{Li} (2013, May).
\newblock {Voids in modified gravity: excursion set predictions}.
\newblock {\em mnras\/}~{\em 431}, 749--766.

\bibitem[\protect\citeauthoryear{{Correa}, {Paz}, {Padilla}, {Ruiz}, {Angulo},
  and {S{\'a}nchez}}{{Correa} et~al.}{2018}]{Correa}
{Correa}, C.~M., D.~J. {Paz}, N.~D. {Padilla}, A.~N. {Ruiz}, R.~E. {Angulo},
  and A.~G. {S{\'a}nchez} (2018, Nov).
\newblock {Non-fiducial cosmological test from geometrical and dynamical
  distortions around voids}.
\newblock {\em arXiv e-prints\/}, arXiv:1811.12251.

\bibitem[\protect\citeauthoryear{{da Cunha}, {Hopkins}, {Colless}, {Taylor},
  {Blake}, {Howlett}, {Magoulas}, {Lucey}, {Lagos}, {Kuehn}, {Gordon}, {Barat},
  {Bian}, {Wolf}, {Cowley}, {White}, {Achitouv}, {Bilicki}, {Bland-Hawthorn},
  {Bolejko}, {Brown}, {Brown}, {Bryant}, {Croom}, {Davis}, {Driver},
  {Filipovic}, {Hinton}, {Johnston-Hollitt}, {Jones}, {Koribalski}, {Kleiner},
  {Lawrence}, {Lorente}, {Mould}, {Owers}, {Pimbblet}, {Tinney}, {Tothill}, and
  {Watson}}{{da Cunha} et~al.}{2017}]{TaipanWP}
{da Cunha}, E., A.~M. {Hopkins}, M.~{Colless}, E.~N. {Taylor}, C.~{Blake},
  C.~{Howlett}, C.~{Magoulas}, J.~R. {Lucey}, C.~{Lagos}, K.~{Kuehn},
  Y.~{Gordon}, D.~{Barat}, F.~{Bian}, C.~{Wolf}, M.~J. {Cowley}, M.~{White},
  I.~{Achitouv}, M.~{Bilicki}, J.~{Bland-Hawthorn}, K.~{Bolejko}, M.~J.~I.
  {Brown}, R.~{Brown}, J.~{Bryant}, S.~{Croom}, T.~M. {Davis}, S.~P. {Driver},
  M.~D. {Filipovic}, S.~R. {Hinton}, M.~{Johnston-Hollitt}, D.~H. {Jones},
  B.~{Koribalski}, D.~{Kleiner}, J.~{Lawrence}, N.~{Lorente}, J.~{Mould}, M.~S.
  {Owers}, K.~{Pimbblet}, C.~G. {Tinney}, N.~F.~H. {Tothill}, and F.~{Watson}
  (2017, June).
\newblock {The Taipan Galaxy Survey: Scientific Goals and Observing Strategy}.
\newblock {\em ArXiv e-prints\/}.

\bibitem[\protect\citeauthoryear{{de la Torre}, {Guzzo}, {Peacock},
  {Branchini}, {Iovino}, {Granett}, {Abbas}, {Adami}, {Arnouts}, {Bel},
  {Bolzonella}, {Bottini}, {Cappi}, {Coupon}, {Cucciati}, {Davidzon}, {De
  Lucia}, {Fritz}, {Franzetti}, {Fumana}, {Garilli}, {Ilbert}, {Krywult}, {Le
  Brun}, {Le F{\`e}vre}, {Maccagni}, {Ma{\l}ek}, {Marulli}, {McCracken},
  {Moscardini}, {Paioro}, {Percival}, {Polletta}, {Pollo}, {Schlagenhaufer},
  {Scodeggio}, {Tasca}, {Tojeiro}, {Vergani}, {Zanichelli}, {Burden}, {Di
  Porto}, {Marchetti}, {Marinoni}, {Mellier}, {Monaco}, {Nichol}, {Phleps},
  {Wolk}, and {Zamorani}}{{de la Torre} et~al.}{2013}]{delaTorre2013}
{de la Torre}, S., L.~{Guzzo}, J.~A. {Peacock}, E.~{Branchini}, A.~{Iovino},
  B.~R. {Granett}, U.~{Abbas}, C.~{Adami}, S.~{Arnouts}, J.~{Bel},
  M.~{Bolzonella}, D.~{Bottini}, A.~{Cappi}, J.~{Coupon}, O.~{Cucciati},
  I.~{Davidzon}, G.~{De Lucia}, A.~{Fritz}, P.~{Franzetti}, M.~{Fumana},
  B.~{Garilli}, O.~{Ilbert}, J.~{Krywult}, V.~{Le Brun}, O.~{Le F{\`e}vre},
  D.~{Maccagni}, K.~{Ma{\l}ek}, F.~{Marulli}, H.~J. {McCracken},
  L.~{Moscardini}, L.~{Paioro}, W.~J. {Percival}, M.~{Polletta}, A.~{Pollo},
  H.~{Schlagenhaufer}, M.~{Scodeggio}, L.~A.~M. {Tasca}, R.~{Tojeiro},
  D.~{Vergani}, A.~{Zanichelli}, A.~{Burden}, C.~{Di Porto}, A.~{Marchetti},
  C.~{Marinoni}, Y.~{Mellier}, P.~{Monaco}, R.~C. {Nichol}, S.~{Phleps},
  M.~{Wolk}, and G.~{Zamorani} (2013, September).
\newblock {The VIMOS Public Extragalactic Redshift Survey (VIPERS) . Galaxy
  clustering and redshift-space distortions at z {$\sim$} 0.8 in the first data
  release}.
\newblock {\em aap\/}~{\em 557}, A54.

\bibitem[\protect\citeauthoryear{{Fisher}}{{Fisher}}{1995}]{Fisher95}
{Fisher}, K.~B. (1995, August).
\newblock {On the Validity of the Streaming Model for the Redshift-Space
  Correlation Function in the Linear Regime}.
\newblock {\em apj\/}~{\em 448}, 494.

\bibitem[\protect\citeauthoryear{{Hamaus}, {Cousinou}, {Pisani}, {Aubert},
  {Escoffier}, and {Weller}}{{Hamaus} et~al.}{2017}]{HamausSDSS2}
{Hamaus}, N., M.-C. {Cousinou}, A.~{Pisani}, M.~{Aubert}, S.~{Escoffier}, and
  J.~{Weller} (2017, May).
\newblock {Multipole analysis of redshift-space distortions around cosmic
  voids}.
\newblock {\em ArXiv e-prints\/}.

\bibitem[\protect\citeauthoryear{{Hamaus}, {Pisani}, {Sutter}, {Lavaux},
  {Escoffier}, {Wand elt}, and {Weller}}{{Hamaus} et~al.}{2016}]{HamausSDSS}
{Hamaus}, N., A.~{Pisani}, P.~M. {Sutter}, G.~{Lavaux}, S.~{Escoffier}, B.~D.
  {Wand elt}, and J.~{Weller} (2016, Aug).
\newblock {Constraints on Cosmology and Gravity from the Dynamics of Voids}.
\newblock {\em prl\/}~{\em 117}, 091302.

\bibitem[\protect\citeauthoryear{{Hamaus}, {Sutter}, {Lavaux}, and
  {Wandelt}}{{Hamaus} et~al.}{2015}]{Hamaustheo}
{Hamaus}, N., P.~M. {Sutter}, G.~{Lavaux}, and B.~D. {Wandelt} (2015,
  November).
\newblock {Probing cosmology and gravity with redshift-space distortions around
  voids}.
\newblock {\em jcap\/}~{\em 11}, 036.

\bibitem[\protect\citeauthoryear{{Hamilton}}{{Hamilton}}{1998}]{Hamilton1992}
{Hamilton}, A.~J.~S. (1998, January).
\newblock {Linear Redshift Distortions: a Review}.
\newblock In D.~{Hamilton} (Ed.), {\em The Evolving Universe}, Volume 231 of
  {\em Astrophysics and Space Science Library}, pp.\  185.

\bibitem[\protect\citeauthoryear{{Hawken}, {Granett}, {Iovino}, {Guzzo},
  {Peacock}, {de la Torre}, {Garilli}, {Bolzonella}, {Scodeggio}, {Abbas},
  {Adami}, {Bottini}, {Cappi}, {Cucciati}, {Davidzon}, {Fritz}, {Franzetti},
  {Krywult}, {Le Brun}, {Le Fevre}, {Maccagni}, {Ma{\l}ek}, {Marulli},
  {Polletta}, {Pollo}, {Tasca}, {Tojeiro}, {Vergani}, {Zanichelli}, {Arnouts},
  {Bel}, {Branchini}, {De Lucia}, {Ilbert}, {Moscardini}, and
  {Percival}}{{Hawken} et~al.}{2016}]{AH_Viper2016}
{Hawken}, A.~J., B.~R. {Granett}, A.~{Iovino}, L.~{Guzzo}, J.~A. {Peacock},
  S.~{de la Torre}, B.~{Garilli}, M.~{Bolzonella}, M.~{Scodeggio}, U.~{Abbas},
  C.~{Adami}, D.~{Bottini}, A.~{Cappi}, O.~{Cucciati}, I.~{Davidzon},
  A.~{Fritz}, P.~{Franzetti}, J.~{Krywult}, V.~{Le Brun}, O.~{Le Fevre},
  D.~{Maccagni}, K.~{Ma{\l}ek}, F.~{Marulli}, M.~{Polletta}, A.~{Pollo},
  L.~A.~.~M. {Tasca}, R.~{Tojeiro}, D.~{Vergani}, A.~{Zanichelli},
  S.~{Arnouts}, J.~{Bel}, E.~{Branchini}, G.~{De Lucia}, O.~{Ilbert},
  L.~{Moscardini}, and W.~J. {Percival} (2016, November).
\newblock {The VIMOS Public Extragalactic Redshift Survey: Measuring the growth
  rate of structure around cosmic voids}.
\newblock {\em ArXiv e-prints\/}.

\bibitem[\protect\citeauthoryear{{Hu} and {Sawicki}}{{Hu} and
  {Sawicki}}{2007}]{Hu_Sawicki_2007}
{Hu}, W. and I.~{Sawicki} (2007, September).
\newblock {Models of f(R) cosmic acceleration that evade solar system tests}.
\newblock {\em prd\/}~{\em 76\/}(6), 064004.

\bibitem[\protect\citeauthoryear{{Huterer}, {Kirkby}, {Bean}, {Connolly},
  {Dawson}, {Dodelson}, {Evrard}, {Jain}, {Jarvis}, {Linder}, {Mandelbaum},
  {May}, {Raccanelli}, {Reid}, {Rozo}, {Schmidt}, {Sehgal}, {Slosar}, {van
  Engelen}, {Wu}, and {Zhao}}{{Huterer} et~al.}{2015}]{Huterer}
{Huterer}, D., D.~{Kirkby}, R.~{Bean}, A.~{Connolly}, K.~{Dawson},
  S.~{Dodelson}, A.~{Evrard}, B.~{Jain}, M.~{Jarvis}, E.~{Linder},
  R.~{Mandelbaum}, M.~{May}, A.~{Raccanelli}, B.~{Reid}, E.~{Rozo},
  F.~{Schmidt}, N.~{Sehgal}, A.~{Slosar}, A.~{van Engelen}, H.-Y. {Wu}, and
  G.~{Zhao} (2015, Mar).
\newblock {Growth of cosmic structure: Probing dark energy beyond expansion}.
\newblock {\em Astroparticle Physics\/}~{\em 63}, 23--41.

\bibitem[\protect\citeauthoryear{{Jones}, {Saunders}, {Colless}, {Read},
  {Parker}, {Watson}, {Campbell}, {Burkey}, {Mauch}, {Moore}, {Hartley},
  {Cass}, {James}, {Russell}, {Fiegert}, {Dawe}, {Huchra}, {Jarrett}, {Lahav},
  {Lucey}, {Mamon}, {Proust}, {Sadler}, and {Wakamatsu}}{{Jones}
  et~al.}{2004}]{Jones2004}
{Jones}, D.~H., W.~{Saunders}, M.~{Colless}, M.~A. {Read}, Q.~A. {Parker},
  F.~G. {Watson}, L.~A. {Campbell}, D.~{Burkey}, T.~{Mauch}, L.~{Moore},
  M.~{Hartley}, P.~{Cass}, D.~{James}, K.~{Russell}, K.~{Fiegert}, J.~{Dawe},
  J.~{Huchra}, T.~{Jarrett}, O.~{Lahav}, J.~{Lucey}, G.~A. {Mamon},
  D.~{Proust}, E.~M. {Sadler}, and K.-i. {Wakamatsu} (2004, December).
\newblock {The 6dF Galaxy Survey: samples, observational techniques and the
  first data release}.
\newblock {\em mnras\/}~{\em 355}, 747--763.

\bibitem[\protect\citeauthoryear{{Kaiser}}{{Kaiser}}{1987}]{Kaiser1987}
{Kaiser}, N. (1987, July).
\newblock {Clustering in real space and in redshift space}.
\newblock {\em mnras\/}~{\em 227}, 1--21.

\bibitem[\protect\citeauthoryear{{Khoury} and {Weltman}}{{Khoury} and
  {Weltman}}{2004}]{Khoury2004}
{Khoury}, J. and A.~{Weltman} (2004, February).
\newblock {Chameleon cosmology}.
\newblock {\em \prd\/}~{\em 69\/}(4), 044026.

\bibitem[\protect\citeauthoryear{{Kopp}, {Uhlemann}, and {Achitouv}}{{Kopp}
  et~al.}{2016}]{KoppUA}
{Kopp}, M., C.~{Uhlemann}, and I.~{Achitouv} (2016, June).
\newblock {Choose to smooth: Gaussian streaming with the truncated Zel'dovich
  approximation}.
\newblock {\em ArXiv e-prints\/}.

\bibitem[\protect\citeauthoryear{{Kreisch}, {Pisani}, {Carbone}, {Liu},
  {Hawken}, {Massara}, {Spergel}, and {Wand elt}}{{Kreisch}
  et~al.}{2018}]{Kreisch}
{Kreisch}, C.~D., A.~{Pisani}, C.~{Carbone}, J.~{Liu}, A.~J. {Hawken},
  E.~{Massara}, D.~N. {Spergel}, and B.~D. {Wand elt} (2018, Aug).
\newblock {Massive Neutrinos Leave Fingerprints on Cosmic Voids}.
\newblock {\em arXiv e-prints\/}, arXiv:1808.07464.

\bibitem[\protect\citeauthoryear{{Linder}}{{Linder}}{2005}]{Linder}
{Linder}, E.~V. (2005, Aug).
\newblock {Cosmic growth history and expansion history}.
\newblock {\em prd\/}~{\em 72}, 043529.

\bibitem[\protect\citeauthoryear{{Nadathur} and {Percival}}{{Nadathur} and
  {Percival}}{2017}]{Nadathur17}
{Nadathur}, S. and W.~J. {Percival} (2017, December).
\newblock {An accurate linear model for redshift space distortions in the
  void-galaxy correlation function}.
\newblock {\em ArXiv e-prints\/}.

\bibitem[\protect\citeauthoryear{{Paillas}, {Cautun}, {Li}, {Cai}, {Padilla},
  {Armijo}, and {Bose}}{{Paillas} et~al.}{2019}]{Paillas}
{Paillas}, E., M.~{Cautun}, B.~{Li}, Y.-C. {Cai}, N.~{Padilla}, J.~{Armijo},
  and S.~{Bose} (2019, Mar).
\newblock {The Santiago-Harvard-Edinburgh-Durham void comparison II: unveiling
  the Vainshtein screening using weak lensing}.
\newblock {\em mnras\/}~{\em 484}, 1149--1165.

\bibitem[\protect\citeauthoryear{{Peacock}, {Cole}, {Norberg}, {Baugh},
  {Bland-Hawthorn}, {Bridges}, {Cannon}, {Colless}, {Collins}, {Couch},
  {Dalton}, {Deeley}, {De Propris}, {Driver}, {Efstathiou}, {Ellis}, {Frenk},
  {Glazebrook}, {Jackson}, {Lahav}, {Lewis}, {Lumsden}, {Maddox}, {Percival},
  {Peterson}, {Price}, {Sutherland}, and {Taylor}}{{Peacock}
  et~al.}{2001}]{Peacock2001}
{Peacock}, J.~A., S.~{Cole}, P.~{Norberg}, C.~M. {Baugh}, J.~{Bland-Hawthorn},
  T.~{Bridges}, R.~D. {Cannon}, M.~{Colless}, C.~{Collins}, W.~{Couch},
  G.~{Dalton}, K.~{Deeley}, R.~{De Propris}, S.~P. {Driver}, G.~{Efstathiou},
  R.~S. {Ellis}, C.~S. {Frenk}, K.~{Glazebrook}, C.~{Jackson}, O.~{Lahav},
  I.~{Lewis}, S.~{Lumsden}, S.~{Maddox}, W.~J. {Percival}, B.~A. {Peterson},
  I.~{Price}, W.~{Sutherland}, and K.~{Taylor} (2001, March).
\newblock {A measurement of the cosmological mass density from clustering in
  the 2dF Galaxy Redshift Survey}.
\newblock {\em nat\/}~{\em 410}, 169--173.

\bibitem[\protect\citeauthoryear{{Peebles}}{{Peebles}}{1993}]{Peebles}
{Peebles}, P.~J.~E. (1993).
\newblock {\em {Principles of Physical Cosmology}}.

\bibitem[\protect\citeauthoryear{{Planck Collaboration}, {Ade}, {Aghanim},
  {Arnaud}, {Ashdown}, {Aumont}, {Baccigalupi}, {Banday}, {Barreiro},
  {Bartlett}, and et~al.}{{Planck Collaboration} et~al.}{2016}]{Planck}
{Planck Collaboration}, P.~A.~R. {Ade}, N.~{Aghanim}, M.~{Arnaud},
  M.~{Ashdown}, J.~{Aumont}, C.~{Baccigalupi}, A.~J. {Banday}, R.~B.
  {Barreiro}, J.~G. {Bartlett}, and et~al. (2016, September).
\newblock {Planck 2015 results. XIII. Cosmological parameters}.
\newblock {\em aap\/}~{\em 594}, A13.

\bibitem[\protect\citeauthoryear{{Pollina}, {Hamaus}, {Dolag}, {Weller},
  {Baldi}, and {Moscardini}}{{Pollina} et~al.}{2017}]{Pollina16}
{Pollina}, G., N.~{Hamaus}, K.~{Dolag}, J.~{Weller}, M.~{Baldi}, and
  L.~{Moscardini} (2017, July).
\newblock {On the linearity of tracer bias around voids}.
\newblock {\em mnras\/}~{\em 469}, 787--799.

\bibitem[\protect\citeauthoryear{{Pollina}, {Hamaus}, {Paech}, {Dolag},
  {Weller}, {S{\'a}nchez}, {Rykoff}, {Jain}, {Abbott}, {Allam}, {Avila},
  {Bernstein}, {Bertin}, {Brooks}, {Burke}, {Carnero Rosell}, {Carrasco Kind},
  {Carretero}, {Cunha}, {D'Andrea}, {da Costa}, {De Vicente}, {DePoy}, {Desai},
  {Diehl}, {Doel}, {Evrard}, {Flaugher}, {Fosalba}, {Frieman},
  {Garc{\'{\i}}a-Bellido}, {Gerdes}, {Giannantonio}, {Gruen}, {Gschwend},
  {Gutierrez}, {Hartley}, {Hollowood}, {Honscheid}, {Hoyle}, {James},
  {Jeltema}, {Kuehn}, {Kuropatkin}, {Lima}, {March}, {Marshall}, {Melchior},
  {Menanteau}, {Miquel}, {Plazas}, {Romer}, {Sanchez}, {Scarpine}, {Schindler},
  {Schubnell}, {Sevilla-Noarbe}, {Smith}, {Soares-Santos}, {Sobreira},
  {Suchyta}, {Tarle}, {Walker}, and {Wester}}{{Pollina}
  et~al.}{2018}]{PollinaDES}
{Pollina}, G., N.~{Hamaus}, K.~{Paech}, K.~{Dolag}, J.~{Weller},
  C.~{S{\'a}nchez}, E.~S. {Rykoff}, B.~{Jain}, T.~M.~C. {Abbott}, S.~{Allam},
  S.~{Avila}, R.~A. {Bernstein}, E.~{Bertin}, D.~{Brooks}, D.~L. {Burke},
  A.~{Carnero Rosell}, M.~{Carrasco Kind}, J.~{Carretero}, C.~E. {Cunha}, C.~B.
  {D'Andrea}, L.~N. {da Costa}, J.~{De Vicente}, D.~L. {DePoy}, S.~{Desai},
  H.~T. {Diehl}, P.~{Doel}, A.~E. {Evrard}, B.~{Flaugher}, P.~{Fosalba},
  J.~{Frieman}, J.~{Garc{\'{\i}}a-Bellido}, D.~W. {Gerdes}, T.~{Giannantonio},
  D.~{Gruen}, J.~{Gschwend}, G.~{Gutierrez}, W.~G. {Hartley}, D.~L.
  {Hollowood}, K.~{Honscheid}, B.~{Hoyle}, D.~J. {James}, T.~{Jeltema},
  K.~{Kuehn}, N.~{Kuropatkin}, M.~{Lima}, M.~{March}, J.~L. {Marshall},
  P.~{Melchior}, F.~{Menanteau}, R.~{Miquel}, A.~A. {Plazas}, A.~K. {Romer},
  E.~{Sanchez}, V.~{Scarpine}, R.~{Schindler}, M.~{Schubnell},
  I.~{Sevilla-Noarbe}, M.~{Smith}, M.~{Soares-Santos}, F.~{Sobreira},
  E.~{Suchyta}, G.~{Tarle}, A.~R. {Walker}, and W.~{Wester} (2018, June).
\newblock {On the relative bias of void tracers in the Dark Energy Survey}.
\newblock {\em arXiv e-prints\/}.

\bibitem[\protect\citeauthoryear{{Reid}, {Samushia}, {White}, {Percival},
  {Manera}, {Padmanabhan}, {Ross}, {S{\'a}nchez}, {Bailey}, {Bizyaev},
  {Bolton}, {Brewington}, {Brinkmann}, {Brownstein}, {Cuesta}, {Eisenstein},
  {Gunn}, {Honscheid}, {Malanushenko}, {Malanushenko}, {Maraston}, {McBride},
  {Muna}, {Nichol}, {Oravetz}, {Pan}, {de Putter}, {Roe}, {Ross}, {Schlegel},
  {Schneider}, {Seo}, {Shelden}, {Sheldon}, {Simmons}, {Skibba}, {Snedden},
  {Swanson}, {Thomas}, {Tinker}, {Tojeiro}, {Verde}, {Wake}, {Weaver},
  {Weinberg}, {Zehavi}, and {Zhao}}{{Reid} et~al.}{2012}]{Reid2012}
{Reid}, B.~A., L.~{Samushia}, M.~{White}, W.~J. {Percival}, M.~{Manera},
  N.~{Padmanabhan}, A.~J. {Ross}, A.~G. {S{\'a}nchez}, S.~{Bailey},
  D.~{Bizyaev}, A.~S. {Bolton}, H.~{Brewington}, J.~{Brinkmann}, J.~R.
  {Brownstein}, A.~J. {Cuesta}, D.~J. {Eisenstein}, J.~E. {Gunn},
  K.~{Honscheid}, E.~{Malanushenko}, V.~{Malanushenko}, C.~{Maraston}, C.~K.
  {McBride}, D.~{Muna}, R.~C. {Nichol}, D.~{Oravetz}, K.~{Pan}, R.~{de Putter},
  N.~A. {Roe}, N.~P. {Ross}, D.~J. {Schlegel}, D.~P. {Schneider}, H.-J. {Seo},
  A.~{Shelden}, E.~S. {Sheldon}, A.~{Simmons}, R.~A. {Skibba}, S.~{Snedden},
  M.~E.~C. {Swanson}, D.~{Thomas}, J.~{Tinker}, R.~{Tojeiro}, L.~{Verde}, D.~A.
  {Wake}, B.~A. {Weaver}, D.~H. {Weinberg}, I.~{Zehavi}, and G.-B. {Zhao}
  (2012, November).
\newblock {The clustering of galaxies in the SDSS-III Baryon Oscillation
  Spectroscopic Survey: measurements of the growth of structure and expansion
  rate at z = 0.57 from anisotropic clustering}.
\newblock {\em mnras\/}~{\em 426}, 2719--2737.

\bibitem[\protect\citeauthoryear{{Sutter}, {Lavaux}, {Hamaus}, {Pisani},
  {Wandelt}, {Warren}, {Villaescusa-Navarro}, {Zivick}, {Mao}, and
  {Thompson}}{{Sutter} et~al.}{2015}]{VIDE}
{Sutter}, P.~M., G.~{Lavaux}, N.~{Hamaus}, A.~{Pisani}, B.~D. {Wandelt},
  M.~{Warren}, F.~{Villaescusa-Navarro}, P.~{Zivick}, Q.~{Mao}, and B.~B.
  {Thompson} (2015, Mar).
\newblock {VIDE: The Void IDentification and Examination toolkit}.
\newblock {\em Astronomy and Computing\/}~{\em 9}, 1--9.

\bibitem[\protect\citeauthoryear{{Sutter}, {Lavaux}, {Hamaus}, {Wandelt},
  {Weinberg}, and {Warren}}{{Sutter} et~al.}{2014}]{Sutter13}
{Sutter}, P.~M., G.~{Lavaux}, N.~{Hamaus}, B.~D. {Wandelt}, D.~H. {Weinberg},
  and M.~S. {Warren} (2014, July).
\newblock {Sparse sampling, galaxy bias, and voids}.
\newblock {\em mnras\/}~{\em 442}, 462--471.

\bibitem[\protect\citeauthoryear{{Tegmark}, {Eisenstein}, {Strauss},
  {Weinberg}, {Blanton}, {Frieman}, {Fukugita}, {Gunn}, {Hamilton}, {Knapp},
  {Nichol}, {Ostriker}, {Padmanabhan}, {Percival}, {Schlegel}, {Schneider},
  {Scoccimarro}, {Seljak}, {Seo}, {Swanson}, {Szalay}, {Vogeley}, {Yoo},
  {Zehavi}, {Abazajian}, {Anderson}, {Annis}, {Bahcall}, {Bassett}, {Berlind},
  {Brinkmann}, {Budavari}, {Castander}, {Connolly}, {Csabai}, {Doi},
  {Finkbeiner}, {Gillespie}, {Glazebrook}, {Hennessy}, {Hogg}, {Ivezi{\'c}},
  {Jain}, {Johnston}, {Kent}, {Lamb}, {Lee}, {Lin}, {Loveday}, {Lupton},
  {Munn}, {Pan}, {Park}, {Peoples}, {Pier}, {Pope}, {Richmond}, {Rockosi},
  {Scranton}, {Sheth}, {Stebbins}, {Stoughton}, {Szapudi}, {Tucker}, {vanden
  Berk}, {Yanny}, and {York}}{{Tegmark} et~al.}{2006}]{Tegmark2006}
{Tegmark}, M., D.~J. {Eisenstein}, M.~A. {Strauss}, D.~H. {Weinberg}, M.~R.
  {Blanton}, J.~A. {Frieman}, M.~{Fukugita}, J.~E. {Gunn}, A.~J.~S. {Hamilton},
  G.~R. {Knapp}, R.~C. {Nichol}, J.~P. {Ostriker}, N.~{Padmanabhan}, W.~J.
  {Percival}, D.~J. {Schlegel}, D.~P. {Schneider}, R.~{Scoccimarro},
  U.~{Seljak}, H.-J. {Seo}, M.~{Swanson}, A.~S. {Szalay}, M.~S. {Vogeley},
  J.~{Yoo}, I.~{Zehavi}, K.~{Abazajian}, S.~F. {Anderson}, J.~{Annis}, N.~A.
  {Bahcall}, B.~{Bassett}, A.~{Berlind}, J.~{Brinkmann}, T.~{Budavari},
  F.~{Castander}, A.~{Connolly}, I.~{Csabai}, M.~{Doi}, D.~P. {Finkbeiner},
  B.~{Gillespie}, K.~{Glazebrook}, G.~S. {Hennessy}, D.~W. {Hogg}, {\v
  Z}.~{Ivezi{\'c}}, B.~{Jain}, D.~{Johnston}, S.~{Kent}, D.~Q. {Lamb}, B.~C.
  {Lee}, H.~{Lin}, J.~{Loveday}, R.~H. {Lupton}, J.~A. {Munn}, K.~{Pan},
  C.~{Park}, J.~{Peoples}, J.~R. {Pier}, A.~{Pope}, M.~{Richmond},
  C.~{Rockosi}, R.~{Scranton}, R.~K. {Sheth}, A.~{Stebbins}, C.~{Stoughton},
  I.~{Szapudi}, D.~L. {Tucker}, D.~E. {vanden Berk}, B.~{Yanny}, and D.~G.
  {York} (2006, December).
\newblock {Cosmological constraints from the SDSS luminous red galaxies}.
\newblock {\em prd\/}~{\em 74\/}(12), 123507.

\bibitem[\protect\citeauthoryear{{White}, {Tinker}, and {McBride}}{{White}
  et~al.}{2014}]{QPM}
{White}, M., J.~L. {Tinker}, and C.~K. {McBride} (2014, Jan).
\newblock {Mock galaxy catalogues using the quick particle mesh method}.
\newblock {\em mnras\/}~{\em 437}, 2594--2606.

\bibitem[\protect\citeauthoryear{{Zivick}, {Sutter}, {Wandelt}, {Li}, and
  {Lam}}{{Zivick} et~al.}{2015}]{Zivicketal2015}
{Zivick}, P., P.~M. {Sutter}, B.~D. {Wandelt}, B.~{Li}, and T.~Y. {Lam} (2015,
  August).
\newblock {Using cosmic voids to distinguish f(R) gravity in future galaxy
  surveys}.
\newblock {\em mnras\/}~{\em 451}, 4215--4222.

\end{thebibliography}

\end{document}